\newif\ifasanka

\documentclass[final,5p,times,twocolumn]{elsarticle}

\usepackage{graphicx}
\usepackage{subfig}
\usepackage{flushend} 
\usepackage{ccicons}
\usepackage{makerobust} 
\MakeRobustCommand\rotatebox 
\graphicspath{{./pdf/}{./jpeg/}{./images/}}
\DeclareGraphicsExtensions{.pdf,.jpeg,.jpg,.png,}
\DeclareGraphicsExtensions{.pdf,.jpeg,.jpg,.png} 
\usepackage{amssymb}
\usepackage{pifont}

\usepackage{pseudocode}
\usepackage{verbatim}
\usepackage{sverb}

\usepackage{amsmath}
\usepackage[ruled]{algorithm2e}
\usepackage[noend]{algpseudocode}

\usepackage{caption} 

\usepackage{textcomp}

\usepackage{upgreek}

\usepackage{url}
\usepackage{float}

\usepackage{amsmath}

\usepackage[T1]{fontenc}

\usepackage{array}

\usepackage{color}

\usepackage{units}
\usepackage{multirow}
\usepackage{soul}
\usepackage{color}
	\definecolor{celadon}{rgb}{0.67, 0.88, 0.69}
  \definecolor{flamingopink}{rgb}{0.99, 0.56, 0.67}

\usepackage{makecell}

\usepackage{enumitem}

\usepackage[table,xcdraw]{xcolor}

\biboptions{numbers}

\usepackage{xspace}

\begin{document}

\begin{frontmatter}

\title{Ensuring Cross-Device Portability of Electromagnetic Side-Channel Analysis\\for Digital Forensics}


 \author[add1,add3]{Lojenaa Navanesan}
 \ead{lojenaa@vau.ac.lk}
 \cortext[mycorrespondingauthor]{Corresponding author}
 \author[add2]{Nhien-An Le-Khac}
 \ead{an.lekhac@ucd.ie}
 \author[add2]{Mark Scanlon}
 \ead{mark.scanlon@ucd.ie}
 \author[add3]{Kasun De Zoysa}
 \ead{kasun@ucsc.cmb.ac.lk}
 \author[add3]{Asanka P. Sayakkara}
 \ead{asa@ucsc.cmb.ac.lk}
 \address[add1]{University of Vavuniya, Sri Lanka}
 \address[add3]{University of Colombo School of Computing (UCSC), Sri Lanka}
 \address[add2]{University College Dublin, Ireland}


\begin{abstract}

Investigation on smart devices has become an essential subdomain in digital forensics.
The inherent diversity and complexity of smart devices pose a challenge to the extraction of evidence without physically tampering with it, which is often a strict requirement in law enforcement and legal proceedings. 
Recently, this has led to the application of non-intrusive Electromagnetic Side-Channel Analysis (EM-SCA) as an emerging approach to extract forensic insights from smart devices. 
EM-SCA for digital forensics is still in its infancy, and has only been tested on a small number of devices so far. 
Most importantly, the question still remains whether Machine Learning (ML) models in EM-SCA are portable across multiple devices to be useful in digital forensics, i.e., \emph{cross-device portability}.
This study experimentally explores this aspect of EM-SCA using a wide set of smart devices.
The experiments using various iPhones and Nordic Semiconductor nRF52-DK devices indicate that the direct application of pre-trained ML models across multiple identical devices does not yield optimal outcomes (under 20\% accuracy in most cases).
Subsequent experiments included collecting distinct samples of EM traces from all the devices to train new ML models with mixed device data; this also fell short of expectations (still below 20\% accuracy). 
This prompted the adoption of transfer learning techniques, which showed promise for cross-model implementations. 
In particular, for the iPhone 13 and nRF52-DK devices, applying transfer learning techniques resulted in achieving the highest accuracy, with accuracy scores of 98\% and 96\%, respectively.
This result makes a significant advancement in the application of EM-SCA to digital forensics by enabling the use of pre-trained models across identical or similar devices. 
\end{abstract}


\end{frontmatter}

\section{Introduction}

The digital forensic investigation of smart devices involves the systematic analysis of electronic evidence within smartphones, tablets, Internet of Things (IoT) devices, and other embedded systems. It aims to uncover, preserve, and interpret digital information, including files, communications, application data, and system logs, for legal or corporate investigative purposes~\cite{ghosh2021systematic}. 
Digital forensics experts utilise specific tools and techniques to identify digital footprints, reproduce events, and provide insights into user behaviour related to criminal activity. 
Such investigations are crucial in legal proceedings to understand the digital interactions of individuals and organisations~\cite{alghamdi2021digital}. 
Digital forensics on smart devices protects individual rights, aids efficient legal procedure, and strengthens cybersecurity efforts, contributing to a safer and more secure digital environment in a world where digital interactions are pervasive~\cite{losavio2019juridical}.

Smart devices provide a number of difficulties that make collecting and analysing digital evidence a challenging task, e.g., encryption, data protection, diversity of devices, cloud-based data, data volume and fragmentation, privacy concerns, rapid technology advancement, anti-forensics techniques, deleted data, legal and jurisdictional hurdles, user authentication, real-time data, and data integrity~\cite{hegarty2014digital, karie2015taxonomy, montasari2019next}. 
Non-invasive techniques in digital forensic investigations refer to methods that do not alter or damage the original digital evidence during the investigation process. These techniques are crucial for preserving the integrity of the evidence and ensuring that it remains admissible in court.

In 2019, Sayakkara et al.~introduced Electromagnetic Side-Channel Analysis (EM-SCA) for the forensic examination of both smartphones and IoT devices~\cite{sayakkara2019leveraging, SAYAKKARA201943}. 
This novel technique revolves around the detection and analysis of electromagnetic signals emitted by the internal components of these devices during their operational activities. 
The EM-SCA technique intensively examines the unintentional electromagnetic radiation emissions generated by components, e.g., processors and memory, in an effort to obtain important insight about the operations, interactions, and potential security flaws of the devices~\cite{asanka2020dissertation}. 
The non-invasive nature of this technique enables the examination of devices without changing their original state, preserving the integrity of the evidence. 
The EM-SCA approach holds significant promise for enhancing digital forensics investigations, as it provides a fresh perspective to uncovering hidden information and potential threats of smartphones and IoT devices, while adhering to strict non-invasive principles~\cite{9324843, sayakkara2021iotemdataset}.

There is an important limitation in EM-SCA that limits its applicability to real-world investigations.
The proposed model is tested using distinct devices, including four different smartphones and four unique IoT devices~\cite{sayakkara2021iotemdataset}. 
However, the existing studies fail to demonstrate that an ML model designed to acquire forensic insights using EM-SCA can be applied to other devices with comparable performance. 
This situation prompts inquiries about the adaptability of a trained ML model across many devices --- even those of the same make and model.

ML models used in EM-SCA are closely linked to the devices that have been used to create the training dataset. It is possible that a trained model may not accurately extract forensic insights from a device, even if the device is of the same make and model. Additionally, Sayakkara et al.'s ML model ~\cite{9324843, sayakkara2021iotemdataset} cannot detect and acquire forensic insights from other devices with shared internal components. In the recent research~\cite{yasarathna2023crossediot}, authors identified some limitations of using EM-SCA in the crossed-IoT devices context, such as device variability, environmental factors, and data collection and processing. They also demonstrated the impact of the multi-core architectures of the processors on the accuracy and reliability of ML models for EM-SCA. Then, they highlighted the possibility of using transfer learning in improving the performance of ML/Deep Leaning models used in analysing EM-SCA data. However authors in~\cite{yasarathna2023crossediot} have not validated their findings with smart devices with complex System-on-Chip (SoC) architectures, in contrast to this work.
In this work, different models of iPhones, representing smartphones, and the Nordic Semiconductor nRF52-DK, representing IoT devices, were chosen to study the cross-device portability of the EM-SCA approach.

In experiments, the number of traces collected for each activity from each device played a crucial role in validating the model's transferability both within the same device and across identical devices. The methodology begins with the utilisation of the Sayakkara et al.'s EM-SCA technique to construct bespoke models for each device. Subsequently, these pre-trained models were directly applied to identical devices to determine the portability of the model across similar-type devices. However, the experimental outcomes did not align with the anticipated results. Consequently, the pre-trained model was tested on different samples of the same device taken at different times. 
Again, this yielded results that did not align with expectations. This also consolidates the findings in~\cite{yasarathna2023crossediot}.

In response to these challenges, the study progressed to the application of transfer learning techniques. 
Specifically, the output layer of the pre-trained model was retrained, resulting in a notable enhancement in accuracy. 
This transformation in the research approach proved to be significant, directing the study towards a more accurate, cross-device portable model implementation. 
Furthermore, this study is confined to conducting experiments solely on identical devices and diverse samples of the same devices. 

This paper makes the following contributions:
\begin{itemize}
    \item Experimentally investigates the behaviour of ML models used in EM-SCA for digital forensics across devices from the same make and model for real-world smart devices of complex SoC architectures.
    \item Examines the impact of using simple ML-based approaches 
    to train models using EM data from a single device to prove that such approaches do not guarantee the same performance of the model on similar or different devices with identical processors.
    \item Demonstrates the effectiveness of transfer learning in addressing cross-device portability of EM-SCA in investigating on smart and IoT devices.
\end{itemize}

The rest of this paper is organised as follows.
The essential information on the background of the field is provided in Section~\ref{status}, followed by the experimental methodology in Section~\ref{methods}.
The experiments and results of the smartphone and IoT device-based studies are described in detail in Section~\ref{experiment}.
Section~\ref{future-direction} provides a detailed discussion on the findings, followed by the conclusion in Section~\ref{conclusion}.

\section{Background} \label{status}

\subsection{Side-Channel Analysis for Digital Forensics} \label{SCA-DF}

Side-channel analysis is a sophisticated technique that exploits unintentional information leakage from electronic devices during their operation. 
This leakage, which includes electromagnetic emissions, power consumption patterns, and timing discrepancies, can provide valuable insights into a device's activities, potentially revealing internal data including cryptographic keys~\cite{standaert2010introduction, buhan2021sok, spreitzer2017systematic}. 
In digital forensics, side-channel analysis can offer a non-invasive approach to inspect devices. 
This method is particularly useful in uncovering information from devices that might be locked or encrypted~\cite{chowdhury2020physical}. 
Despite its advantages, side-channel analysis requires specialised knowledge and tools due to its complexity. 
It has applications in various areas, including cryptography, cybersecurity, and reverse engineering~\cite{lavaud2021whispering, hossain2018dependence}. 

EM-SCA presents a promising opportunity for acquiring forensic insights from electronic devices. 
This approach capitalises on the unintentional electromagnetic radiation emitted during the operation of the devices, which can carry valuable information about the device's activities. 
Sayakkara et al.~demonstrated the applicability of EM-SCA in the context of IoT device forensics, which has the capability to provide forensic insights beyond what conventional methods can achieve~\cite{sayakkara2021iotemdataset}. 
Furthermore, their work highlights how electromagnetic side-channel analysis can be utilised not only for forensic purposes, but also to identify vulnerabilities and potential attack vectors in IoT devices. 
This underscores the versatility of EM-SCA in both offensive and defensive security contexts~\cite{SAYAKKARA201943, SAYAKKARA2020301003, asanka2020dissertation, kar2017enhancing}



\subsection{The Acquisition of EM Side-Channel Radiation} \label{EM-gathering}

The electrical components of computing equipment generate EM radiation as an effect of internal operations. 
Both smartphones and IoT devices have a variety of internal EM emitting components, including processors, RAM, bus lines, network adaptors, video and audio units, etc.
These interior parts are often associated with a System-on-Chip (SoC) that effortlessly generates EM radiation at its system clock frequency. 
This EM radiation can carry crucial information leaked during the operation of internal components. 
Attackers can exploit this leaked information for their own advantage~\cite{sayakkara2021iotemdataset}.

The EM radiation associated with various software behaviours from IoT devices and smartphones has been identified as a raw source for EM-SCA for the acquisition of digital forensic insight~\cite{9324843, sayakkara2021iotemdataset, le2021identifying}. 
Researchers can identify and measure the EM radiation produced during software activities by positioning specialised sensors or probes in close proximity to the device. 
This radiation contains minimal but observable patterns that are related to the particular actions and processes taking place within the components of a device. 
These patterns can be analysed and correlated with the software activities being performed at the time.

Software Defined Radios (SDRs) are a specific kind of wireless hardware equipment that can capture analogue electromagnetic radiation signals and digitise them to be processed with software tools~\cite{asanka2020dissertation, jondral2005software}. 
\emph{HackRF One} SDR is one of such tools that has been used to capture EM radiation from IoT devices and smartphones~\cite{10.1145/3236454.3236512}. 
A H-loop near-field antenna attached to a HackRF One SDR has been used to identify the EM radiation of various software behaviours. 
Usually, the antenna is moved over the target device to get closer to the SoC processor - since it is anticipated that the SoC would leak crucial information loudest about the internal operations of the device.

Collected EM radiation can also provide insights into the software execution on smart devices, revealing information about the type of applications being used, the intensity of processing, and even potential security vulnerabilities. 
This approach can be utilised in digital forensics to reconstruct a timeline of a device's activities, aiding in investigations to understand the sequence of events leading up to a particular situation.

\subsection{Transfer Learning for Digital Forensics} \label{transfer-learning-DF}

Transfer learning has emerged as a prominent trend in the current era of artificial intelligence.
Transfer learning involves extracting insights from one problem domain and applying these insights to address a similar, new problem. 
Transfer learning strategies enable the sharing of knowledge, improving generalisation and overcoming the limitations of isolated learning procedures. 
Pre-trained ML models are particularly recommended for the problem of classification. 
It has the potential to use information from a previously trained model when faced with a new problem, greatly decreasing the time and effort required to build a new model from scratch~\cite{li2021knowledge, pan2009survey}.

Transfer learning is experiencing increased adoption, especially in comparison to supervised learning, in both commercial and research domains~\cite{taylor2009transfer}. The notable advantages of transfer learning over traditional machine learning methods are evident. Traditional learning operates in isolation, requiring substantial data volumes for accurate learning and classification. In contrast, transfer learning leverages knowledge from previously mastered domains, eliminating the need for extensive datasets~\cite{bengio2021machine}. Consequently, transfer learning accelerates processing, conserves memory, reduces space requirements, and saves power. A notable benefit is that one does not need to be a deep learning expert to execute operations; knowledge from analogous situations suffices.

The idea of transfer learning for EM-SCA is a useful and cutting-edge method for digital forensic investigations on smart devices. 
Transfer learning makes use of the vast amounts of data and expertise to improve the precision and effectiveness of forensic investigation by collecting EM traces from specific devices or groups of devices that are similar to those in question~\cite{goundar2023improved, stoyanova2020survey}. 
In this context, transfer learning makes it easier to apply the knowledge learned from one investigation to another, allowing researchers to make use of the patterns and signatures present in EM traces. 
This strategy enables researchers to develop models, algorithms, and approaches that can recognise and interpret EM signals more successfully by leveraging data from known examples or comparable equipment~\cite{pan2009survey}.

\subsection{Cross-Device Portability of Side-Channel Analysis} \label{CDP}

In the context of digital forensic investigation, the importance of a cross-device portable model becomes crucial - especially when smart devices are present at crime scenes. 
Smart devices have become an essential part of human interaction and communication in today's interconnected society, making them potential sources of critical evidence in investigative cases. 
Investigators face a significant hurdle due to the diversity of smart devices, which includes different brands, models, components, and operating systems. 
This problem can be solved by a cross-device portable model that provides a standardised representation of a particular smart device.  
Such a model can be utilised for performing digital forensic analysis on many kinds of smart devices. 
This approach has a multitude of benefits: efficiency, consistency, adaptability, resource optimisation, comprehensive insight, reduced learning curve, and legal credibility.


Performing EM and power side-channel attacks using deep learning models has been the focus of the security community in recent years~\cite{yu2021cross, cao2022pa, zhang2020homogeneous, das2019x, danial2021x, golder2019practical}.
The possibility for adopting distinctive characteristics, such as the applicability of one device's knowledge to another, regardless of whether they share the same manufacturer or belong to completely separate families, is revealed by cross-knowledge, and cross-family side-channel attacks respectively~\cite{thapar2020transca, thapar2021deep}. 
An idea  encompasses the transferability of machine learning models between various types, which means that regardless of the characteristics of each model, the knowledge gained from one model might be useful for another model known as cross-model/cross-domain side-channel attack~\cite{yu2021cross, bird2020cross}. 
In essence, a cross-device portable model can streamline and enhance digital forensic investigations involving smart devices. 
It empowers investigators to efficiently and consistently extract evidence from a diverse range of devices.

\section{Experimental Methodology} \label{methods}

This study is carried out using two different avenues in order to explore cross-device portability among various smart devices in digital forensics investigations. 
The first avenue involves the dedicated collection of a diverse range of smartphones, each segmented based on their unique features and attributes. 
In the second avenue, the study broadens its focus to include IoT devices. 
The selected embedded hardware platform is representative of typical IoT devices.
The subsequent sections provide a detailed explanation of the two avenues.

\subsection{Methodology for EM Data Acquisition} \label{EM-acq}

This study employs the HackRF One SDR, which has a frequency range of 1 MHz to 6 GHz, and a maximum sampling rate of 20 MHz~\cite{sayakkara2021iotemdataset}. Configuration and data processing use the GNU Radio library and its graphical interface, GNU Radio Companion (GRC), for building EM data processing pipeline. The EM radiation under investigation originates from the SoC processor of the Device-Under-Test (DUT), and proximity to the SoC during data acquisition improves signal reception. To achieve this, an RF Explorer near-field H-loop antenna is connected to the HackRF One device for close-proximity data acquisition from the DUT.

Identifying the optimal location for maximum signal reception involves manually moving the near-field antenna while plotting the spectrogram of the received signal at the CPU clock frequency of the DUT. The position where the signal is the strongest is fixed for subsequent EM trace acquisition, forming the dataset. Although existing literature explores tools and algorithms for determining the ideal signal reception location~\cite{danial2020scnifferF}, this study limits the detection of optimum position for each considered DUT to manual observation of signal strength.

The key to obtaining high-quality EM trace data on the HackRF device lies in determining optimal signal amplification values. Setting amplification too low can make it challenging to capture weak EM radiation from a DUT. On the other hand, excessive signal amplification may amplify external noise, resulting in a cluttered EM trace file. The determination of amplification settings involved empirical experimentation with various configurations, considering signal clarity across different devices in the existing work~\cite{sayakkara2021iotemdataset}. Hence, in line with the findings from the previous study, the radio frequency power amplifier (RF), the low noise amplifier (IF), and the variable-gain amplifier (BB) are consistently configured at 14 dB, 40 dB, and 18 dB, respectively, throughout the experiments~\cite{sayakkara2021iotemdataset}.

\subsection{Experiments with Smartphones} \label{Phone-experiment}

In order to perform experiments with smartphones, iPhones were selected as they have a large user base.
The devices were divided according to their version, model, processor type, and architecture. 
Table~\ref{description-of-devices} provides an overview of the specifics of the selected devices. The system clock frequency was used to calibrate GNU Radio Companion (GRC) software to capture EM traces from the location of the SoC of each device. 

\begin{figure}[!t]
	\centering
	\includegraphics[width=0.3\textwidth]{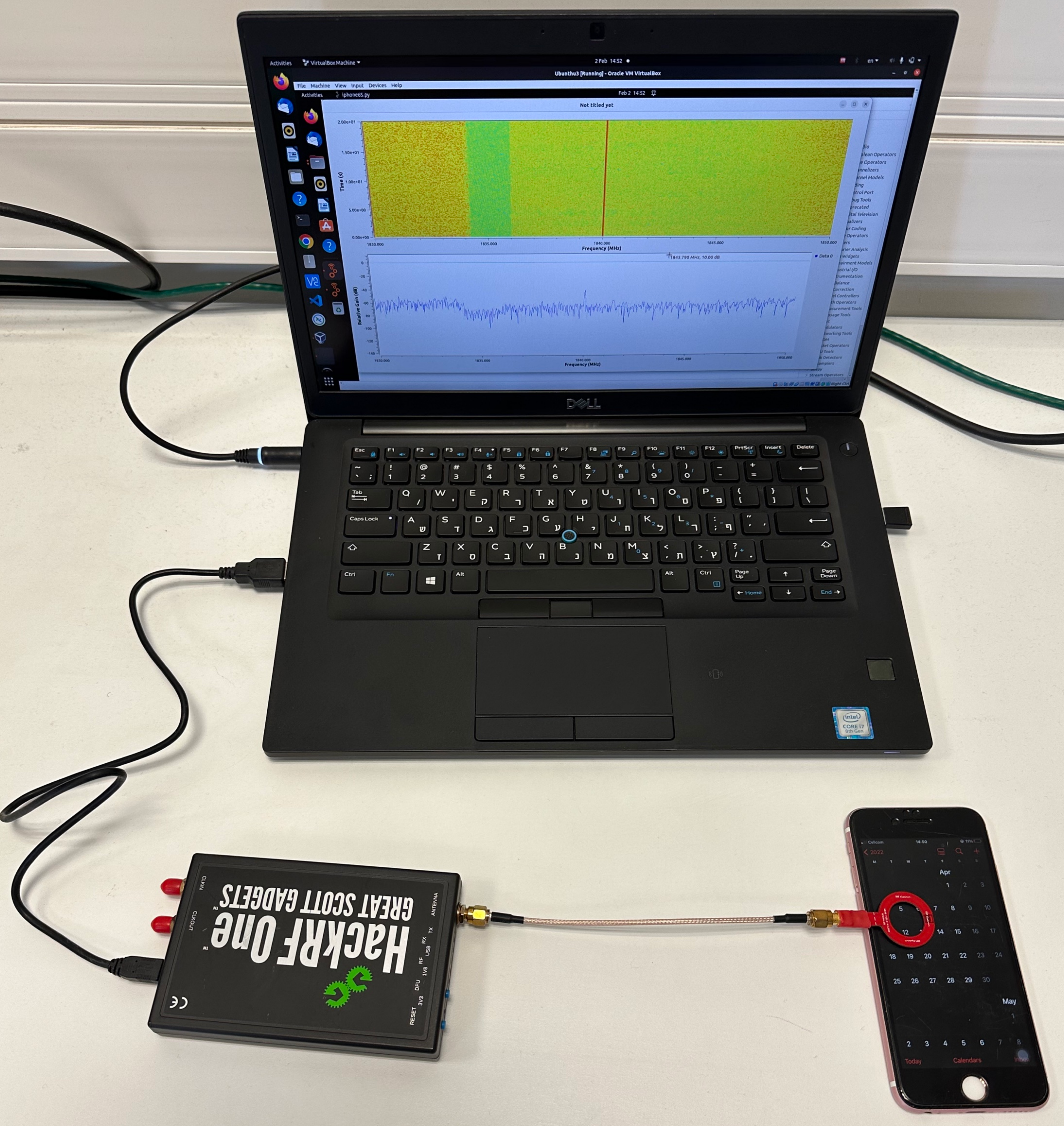}
	\caption{Acquisition of electromagnetic (EM) traces while carrying out various software activities on the iPhone using the HackRF One SDR connected with h-loop near-field antenna and controlled by GNU Radio Companion.}
	\label{iPhone-em-capture}
\end{figure}

\begin{figure}[!t]
	\centering
	\includegraphics[width=0.5\textwidth]{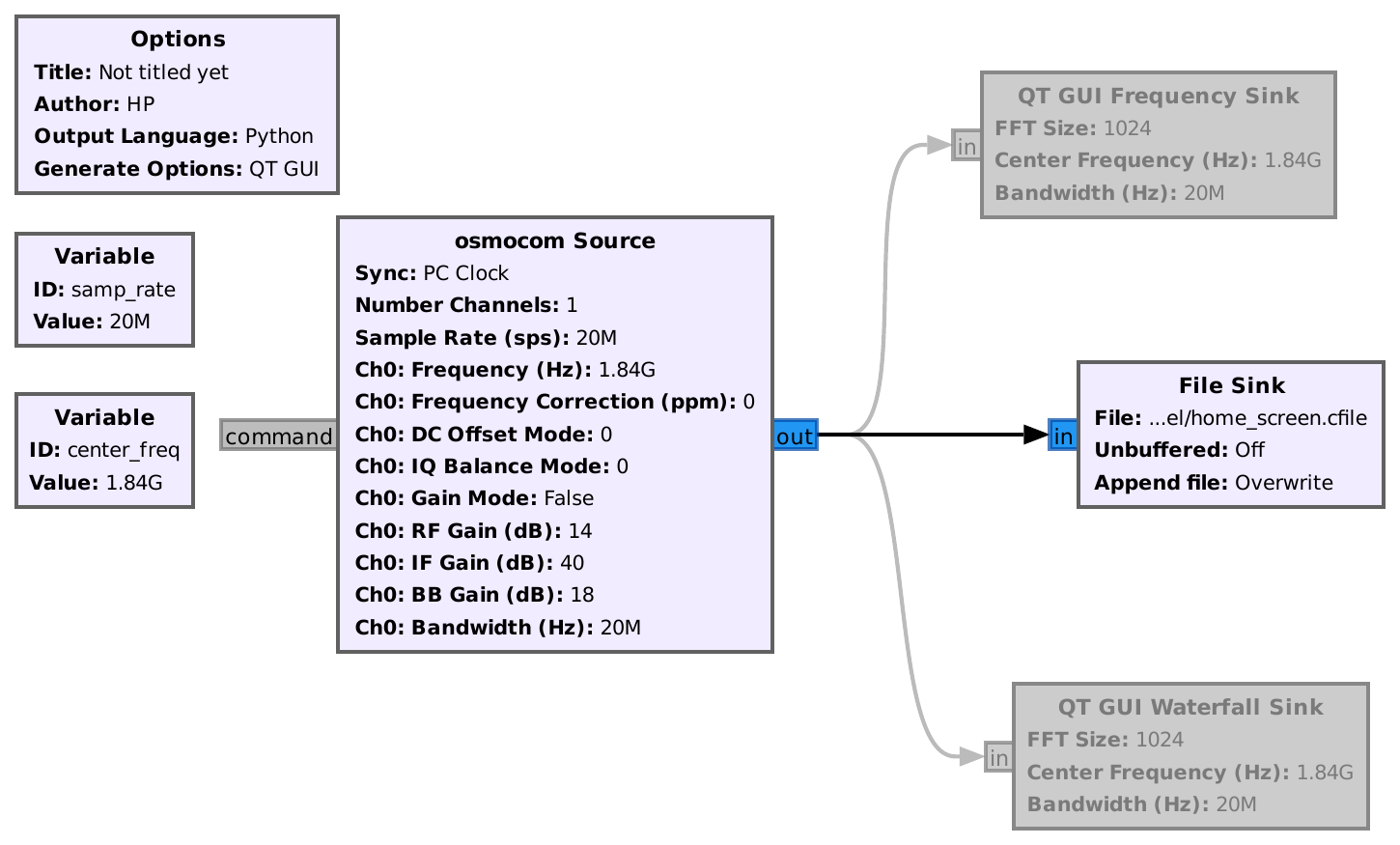}
	\caption{A flow diagram of the smartphone used to set the parameters for acquiring EM traces from each individual smartphone.}
	\label{iPhone-flowgraph}
\end{figure}

The following ten activities were conducted on each of the selected iPhones to observe and collect EM radiation: \textit{calendar-app, camera-photo, camera-video, email-app, gallery-app, home-screen, idle, phone-app, sms-app, and web-browser-app}. 
EM traces of each activity were recorded from each iPhone at its corresponding system clock frequency, as shown in Table~\ref{description-of-devices}. 
This process is described in detail in Section~\ref{EM-gathering}. 
A HackRF One, a computer with the GRC software installed, and a H-loop near-field antenna was used to gather EM traces. 
Figure~\ref{iPhone-em-capture} shows the hardware arrangement of the EM signal acquisition process. 
The GRC software uses a flow graph to construct the parameter configuration for the data gathering, as depicted in Figure~\ref{iPhone-flowgraph}. 
Here, the sample rates are defined as the number of samples per second, and the system clock frequency of each iPhone is allocated to the variable \emph{centre frequency}. 
\emph{Osmocom Source} represents attributes of HackRF One device. 
\emph{Frequency Sink} and \emph{Waterfall Sink} are used to recognise peak signal and the pattern of EM signals at the proper frequency point. 
In addition, \emph{File Sink} is used to store the traces file in the \textit{.cfile} format.

\begin{table}[htp]
\begin{center}
	\caption{Specifications of the targeted devices for capturing EM trace files.}
	\label{description-of-devices}
        \begin{tabular}{|p{12mm}|p{14mm}|l|p{13mm}|p{7mm}|}
	\hline 
	   \textbf{\footnotesize Device} & \textbf{\footnotesize System-on-Chip} & \textbf{\footnotesize Architecture} & \textbf{\footnotesize CPU Frequency (cores)} & \textbf{\footnotesize Device Count}  \\ \hline 
	    \footnotesize iPhone 4S &  \footnotesize Apple A5 &  \footnotesize ARMv7-A &  \footnotesize 1GHz (2) &  \footnotesize 1 \\ \hline 
            \footnotesize iPhone 6S &   \footnotesize Apple A9 &   \footnotesize ARMv8-A &   \footnotesize 1.85GHz (2) &   \footnotesize 1 \\ \hline    
             \footnotesize iPhone 8 &   \footnotesize Apple A11 Bionic &   \footnotesize ARMv8-A &   \footnotesize 2.39GHz (6) &   \footnotesize 1 \\ \hline 
             \footnotesize iPhone 13 &   \footnotesize Apple A15 Bionic &   \footnotesize ARMv8.5-A &   \footnotesize 3.23 GHz (6) &   \footnotesize 3 \\ \hline 
             \footnotesize iPhone 14 Pro &   \footnotesize Apple A16 Bionic &   \footnotesize ARMv8.6-A &   \footnotesize 3.46 GHz (6) &   \footnotesize 1 \\ \hline 
        \end{tabular} 
\end{center}
\end{table}

\begin{figure}[!t]
	\centering
	\includegraphics[width=0.3\textwidth]{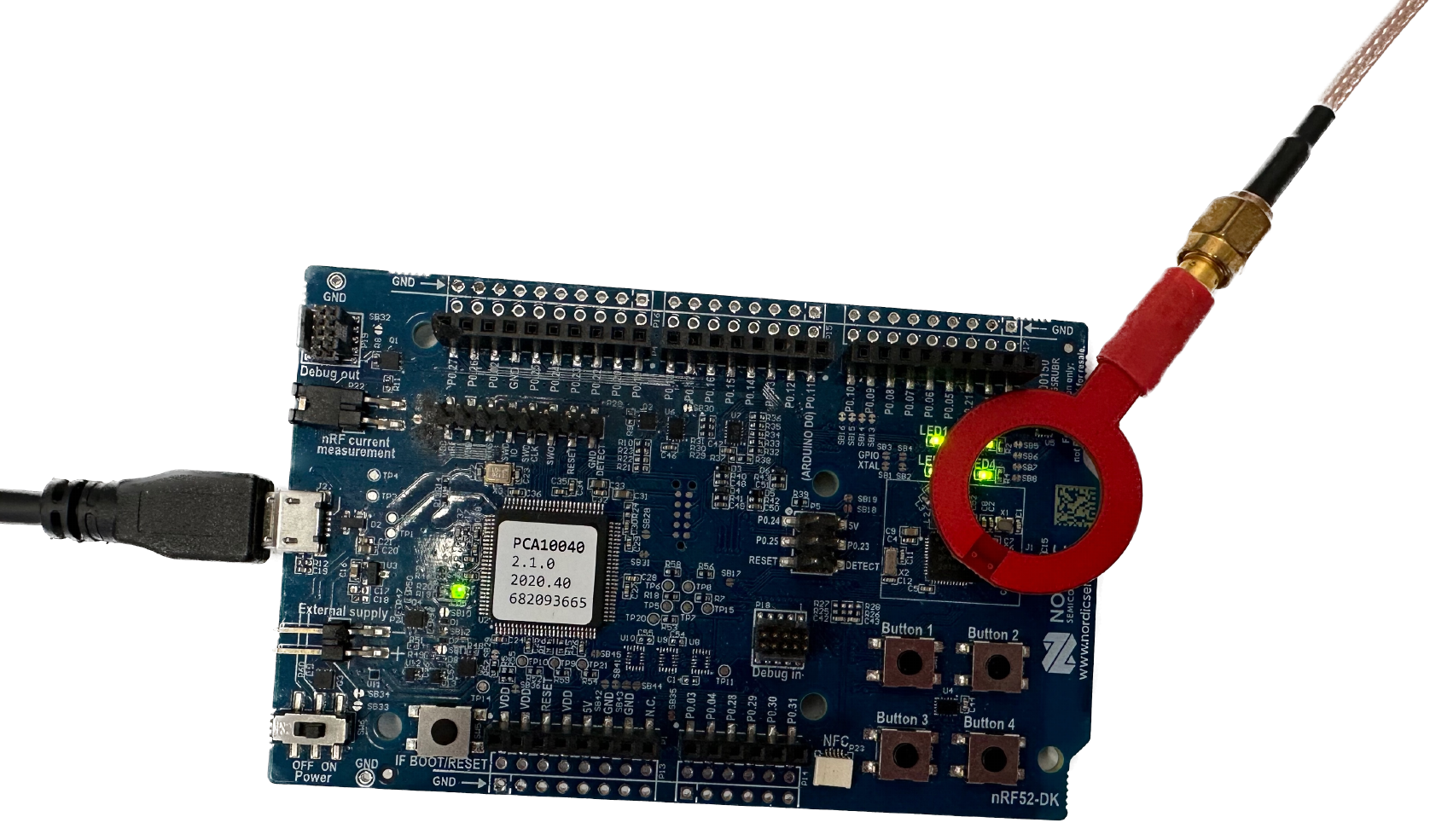}
	\caption{Acquisition of electromagnetic (EM) traces while carrying out various software activities on the Nordic Semiconductor nRF52-DK using the HackRF One SDR connected with h-loop near-field antenna.}
	\label{Nordic-em-capture}
\end{figure}

\subsection{Experiments with IoT}

The Nordic Semiconductor nRF52-DK development kit was selected to represent typical IoT devices.
Two identical nRF52-DK devices were chosen that contain nRF52832 System-on-Chip (SoC) with a maximum system clock frequency of 2.4 GHz.
Eight distinct software activities were selected to capture and evaluate EM traces: \textit{blinky, blinky\_freertos, blinky\_rtc\_freertos, blinky\_systick, led\_softblink, BLINK\_new, IDLE\_new, and Matrix\_multiplication\_new}. 
Each programme is installed into the chip using SEGGER Embedded Studio software. 
The hardware setup for acquiring EM signals from the nRF52-DK devices is shown in Figure~\ref{Nordic-em-capture}.

\begin{figure}[!t]
	\centering
	\includegraphics[width=0.5\textwidth]{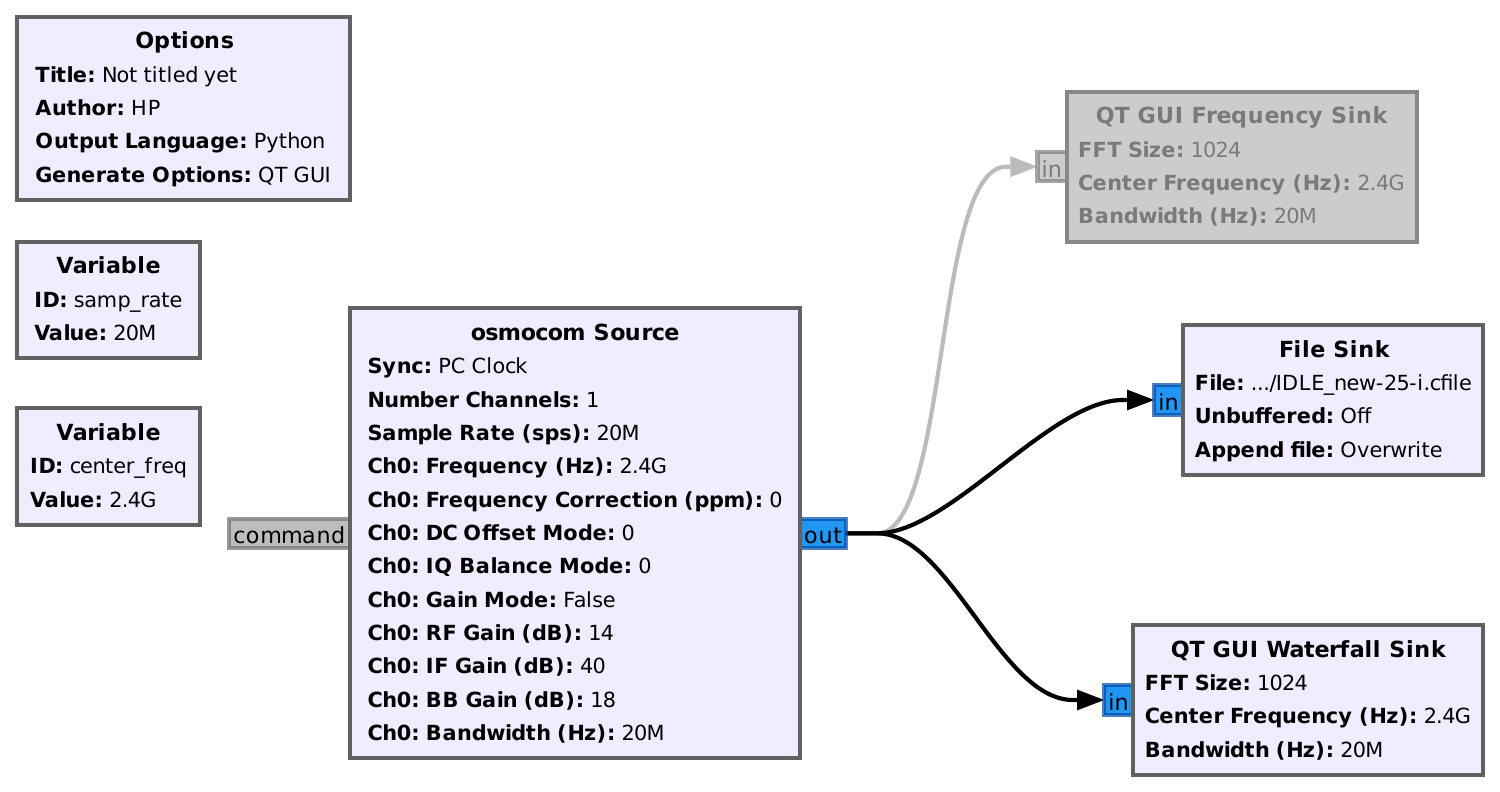}
	\caption{The flow diagram used to set the parameters for acquiring EM traces from each individual Nordic Semiconductor nRF52-DK device.}
	\label{nordic-flowgraph}
\end{figure}

The system clock frequency of the nRF52-DK is set as the center frequency in the GRC utility sofware's flow graph.
Other components, such as Waterfall Sink, Frequency Sinks, and File Sinks, are also included.
Figure~\ref{nordic-flowgraph} depicts the GRC flow graph for collecting EM data from  nRF52-DK devices. 

\begin{figure*}[!t]
	\centering
	\includegraphics[width=0.9\textwidth]{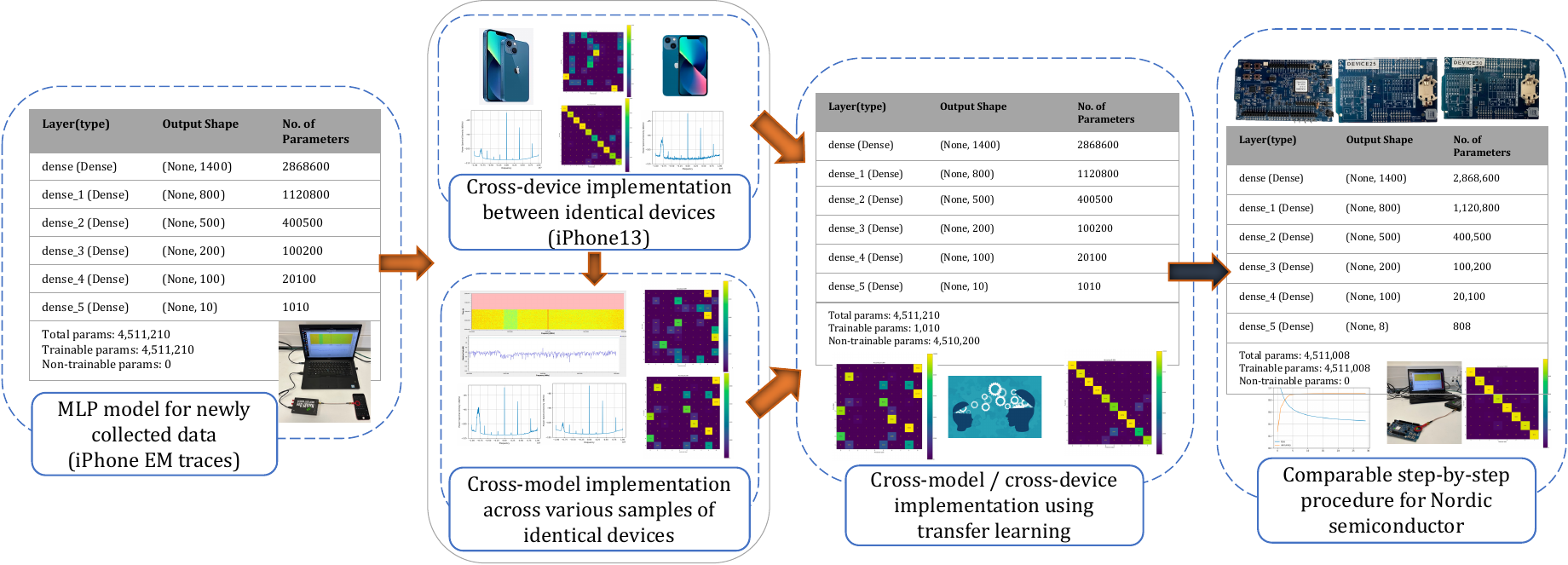}
	\caption{The step-by-step procedure of the experiments involving iPhone and Nordic Semiconductor nRF52-DK devices.}
	\label{roadmap}
\end{figure*}

\subsection{Prepossessing Procedure}

EM radiation was sampled at 20 MHz, resulting in $10$ EM trace files per iPhone and $8$ EM trace files per nRF52-DK device. Each trace file, representing a time-domain signal, underwent Short-Time Fourier Transform (STFT) processing to create frequency-domain windows. In deep learning, these windows served as training instances, labeled with the corresponding smartphone software activities.

The resulting EM datasets for each smartdevice were used to build individual deep learning models for device-specific software activity identification. Certain hyperparameters depended on the EM dataset dimensions, and during hyperparameter tuning, specific STFT operation settings (FFT window size and overlapping samples) were adjusted.

\section{Experiments and Results} \label{experiment}

This section describes the implementation of machine learning-based EM-SCA approach after the EM traces of the iPhones and nRF52-DK devices have been recorded. 
The overall EM-SCA approach for cross-device and cross-model implementation among the selected devices is illustrated in Figure~\ref{roadmap}. 
The data and code used for the experiments of this work are available on a publicly accessible \emph{GitHub} repository~\footnote{https://github.com/Lojenaa/Portability-of-Devices-in-EMSCA.git}. 

\subsection{Experiments with Smartphones} 

\subsubsection{Experiment 1: An ML model per Device} 
\label{ML-for-EMSCA-DF}

Sayakkara et al. used Multi-Layer Perceptron (MLP) machine learning models to identify various software behaviours of smartphones using captured EM traces~\cite{9324843}. 
However, in their work, individual models were developed for each individual device using its corresponding captured EM data. 
To start, this work reproduces the same approach to evaluate its performance on a specific set of iPhones.
Seven EM trace files were collected from the seven different iPhones mentioned in Table~\ref{description-of-devices}. 
Among them, there were 3 iPhone 13 devices, referred to as \emph{iPhone13\_I}, \emph{iPhone13\_II}, and \emph{iPhone13\_III} in the rest of this work.

Using the acquired EM data, a bespoke model was created for each smartphone using their respective datasets.
For this purpose, 10,000 samples from each EM trace file representing a particular software activity were used.
The relevant software activity is considered as the class/label in this instance. 
The structure of the MLP model used for each devices is shown in the Table~\ref{MLP}. 
The input layer of the model consists of 2,048 feature vectors as input shape. There are six intermediate dense layers with a Rectified Linear Units (ReLU) activation function, followed by an output layer with ten nodes that provides the number of classes in each dataset. 
A total of 4,511,210 distinct parameters can be trained on the dataset.

\begin{table}[htp]
\begin{center}
	\caption{The structure of the machine learning model utilising the recently acquired smartphone dataset}
	\label{MLP}
        \begin{tabular}{|l|l|l|l|}
	\hline 
	   \textbf{\footnotesize Layer (type)} & \textbf{\footnotesize Output Shape} & \textbf{\footnotesize No. of Parameters}  \\ \hline 
	   \footnotesize dense (Dense) & \footnotesize (None, 1400) & \footnotesize 2,868,600 \\ \hline 
          \footnotesize dense\_1 (Dense) & \footnotesize (None, 800) & \footnotesize 1,120,800  \\ \hline    
          \footnotesize dense\_2 (Dense) & \footnotesize (None, 500) & \footnotesize 400,500 \\ \hline 
          \footnotesize dense\_3 (Dense) & \footnotesize (None, 200) & \footnotesize 100,200 \\ \hline 
          \footnotesize dense\_4 (Dense) & \footnotesize (None, 100) & \footnotesize 20,100 \\ \hline 
          \footnotesize dense\_5 (Dense) & \footnotesize (None, 10) & \footnotesize 1,010 \\ \hline 
        \end{tabular} 
\end{center}
\end{table}

\begin{figure}[!t]
	\centering
	\includegraphics[width=0.4\textwidth]{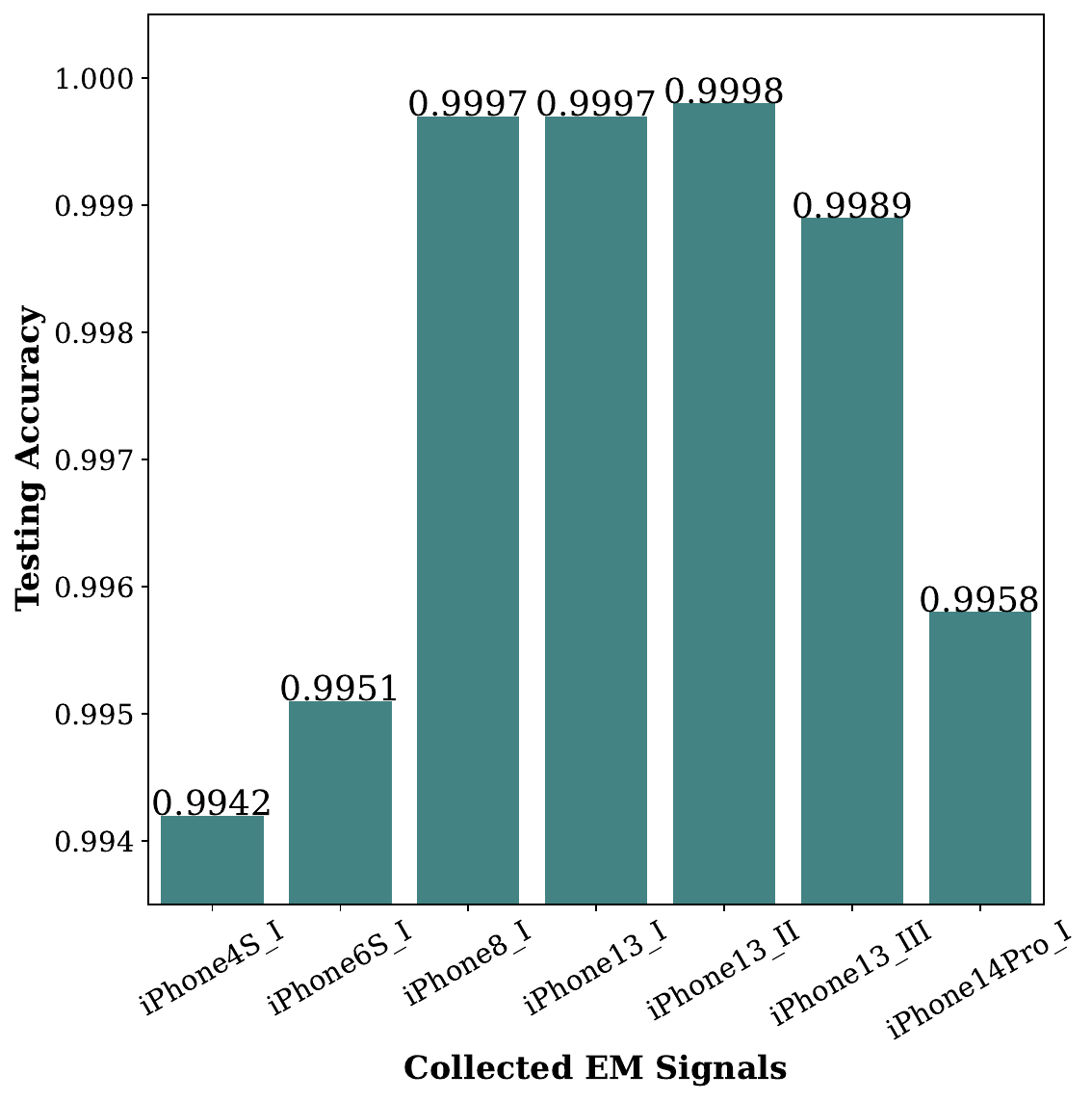}
	\caption{The study applied Sayakkara et al.'s EM-SCA model to assess device accuracy. It utilized specific datasets and EM traces for different iPhone activities, displaying testing accuracy using the MLP machine learning model with activities on the x-axis and accuracy on the y-axis.}
	\label{iPhone13-Accuracy}
\end{figure}

\begin{figure}[!t]
	\centering
	\includegraphics[width=0.4\textwidth]{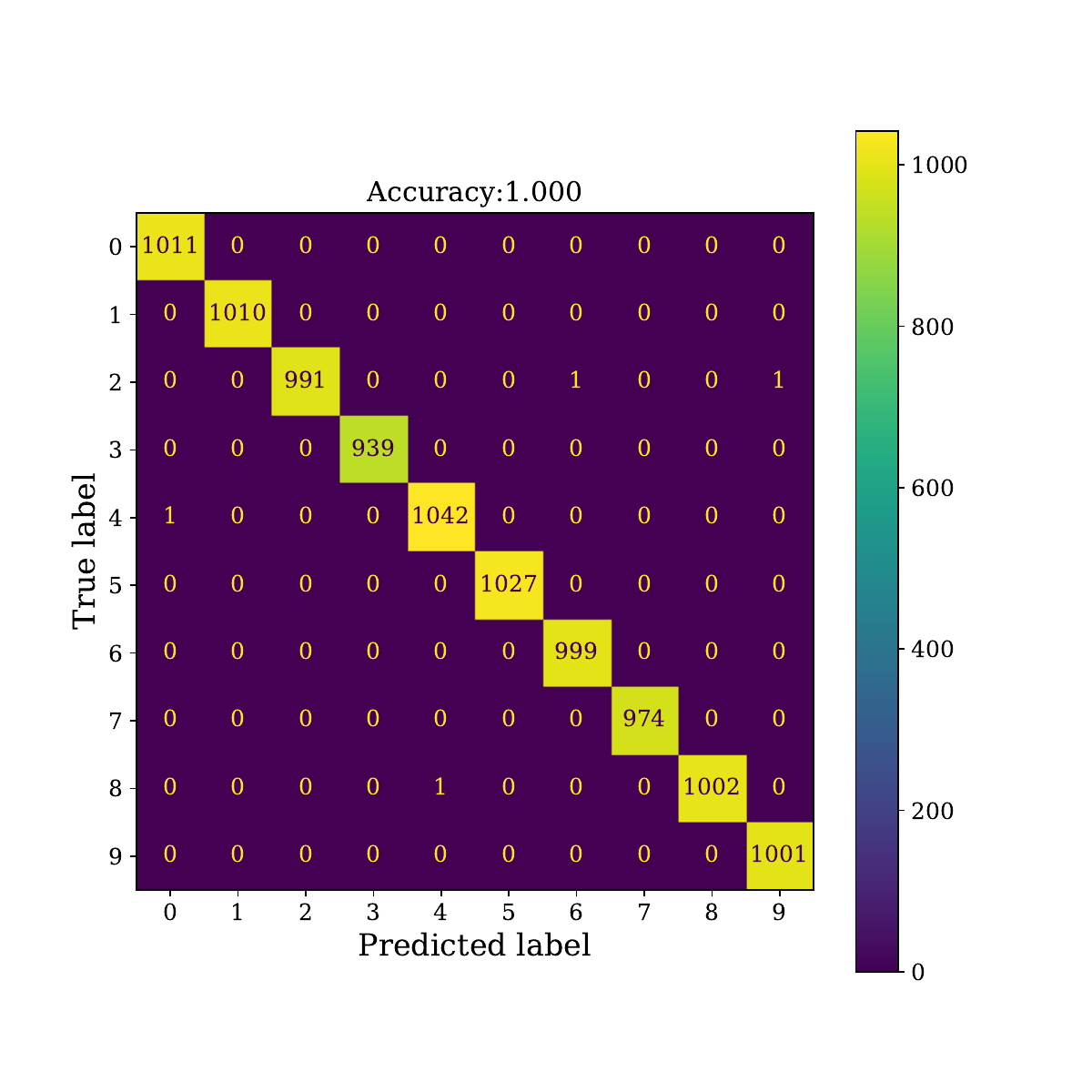}
    	\caption{Confusion matrix of one of the dataset. The implementation of the Sayakkara et al.'s MLP machine learning model while operating ten different iPhone 13 software behaviours.}
	\label{iPhone13-Confusion-Matrix}
\end{figure}

The model performs a 30-epoch training phase. This duration was determined after evaluating the ML model across epochs ranging from 5 to 100, with 30 identified as the optimal number. The training employs the \textit{opt} optimizer and a \textit{sparse categorical cross-entropy} loss function.
Figure~\ref{iPhone13-Accuracy} illustrates the observation of the accuracy of the acquired EM traces over various iPhone types.
The average accuracy of most bespoke models was 99\% when testing on each specific device. 
Additionally, Figure~\ref{iPhone13-Confusion-Matrix} depicts the confusion matrix resulting from validation of one particular iPhone 13 device. 

\subsubsection{Experiment 2: Models across Identical Devices}
\label{cross-device-portability}

The previous experiment demonstrated that a machine learning model trained and tested using the EM data from the same device achieves high accuracy.
However, the question arises whether a trained and tested model on one specific device would perform with similar accuracy when exposed to new testing data acquired from another device of the same make and model.
To explore this aspect, the three ML models were created using the specific traces from each of the three iPhone 13 devices.
Accordingly the model from one device was tested by feeding it samples from the traces of the other two devices.
For instance, the previously saved model for the iPhone 13\_I is immediately fed with to the datasets from the iPhone 13\_II and iPhone 13\_III.
Unfortunately, it was observed that, even if the devices are of the same make and model, the accuracy fell short of expected values; the accuracy was extremely poor (0.1050 and 0.2232 respectively). 
Table~\ref{CDP-iPhone13}'s \textit{Sayakkara et al.'s EM-SCA appraoch} column displays the reaming findings of the direct testing of three iPhone 13 models.


\begin{table*}[t!]
\begin{center}
	\caption{Cross-device portability validation to evaluate the accuracy value by applying Sayakkara et al.'s EM-SCA model on three identical devices of the iPhone 13 device type shown in the first three column under the Testing accuracy and applying transfer learning on the Sayakkara et al.'s EM-SCA approach shown in last three column under the testing accuracy.}
	\label{CDP-iPhone13}
        \begin{tabular}{|l|l|l|l|l|l|l|} \hline 
            \multirow{3}{*}{\textbf{\footnotesize Model Name}} & \multicolumn{6}{c|}{\textbf{\footnotesize Testing Accuracy for each identical iPhone 13}}  \\ \cline{2-7}
                & \multicolumn{3}{c|}{\textbf{\footnotesize Sayakkara et al.'s EM-SCA approach}} & \multicolumn{3}{c|}{\textbf{\footnotesize Transfer learning in Sayakkara et al.'s EM-SCA approach}} \\ \cline{2-7}
             & {\textbf{\footnotesize iPhone13-I}} & {\textbf{\footnotesize iPhone13-II}} & {\textbf{\footnotesize iPhone13-III}} & {\textbf{\footnotesize iPhone13-I}} & {\textbf{\footnotesize iPhone13-II}} & {\textbf{\footnotesize iPhone13-III}}   \\ \hline 
           \footnotesize iPhone13-I-model.h5 & \textbf{\footnotesize 0.9998} & \footnotesize 0.1050 & \footnotesize 0.2232 & \footnotesize - & \footnotesize 0.9559 & \footnotesize 0.7034 \\ \hline 
           \footnotesize iPhone13-II-model.h5 & \footnotesize 0.0938 & \textbf{\footnotesize 0.9998} & \footnotesize 0.1063 & \footnotesize 0.8146 & \footnotesize - & \footnotesize 0.7378 \\ \hline    
           \footnotesize iPhone13-III-model.h5 & \footnotesize 0.1010 & \footnotesize 0.1000 & \textbf{\footnotesize 0.9994} & \footnotesize 0.7000 & \footnotesize 0.8669 & \footnotesize - \\ \hline  
        \end{tabular} 
\end{center}
\end{table*}

\begin{figure}[!t]
	\centering
	\includegraphics[width=0.3\textwidth]{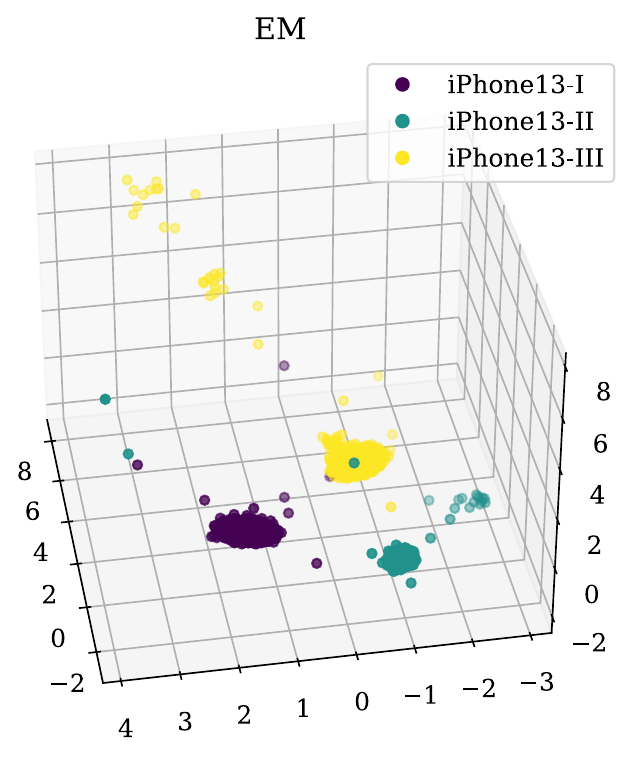}
	\caption{Observation of \textit{idle} activity on three iPhone 13s. A 3D scatter plot of the \textit{idle} activity from three identical iPhone 13 devices  is seen by using the first three components of PCA value from the obtained iPhone 13s EM trace.}
	\label{idle-scatter}
\end{figure}

This experiment shows that existing manner of training models for Sayakkara et al.'s EM-SCA does not adhere to the fundamental concept of cross-device portability. 
Even with three identical iPhone 13s, the model trained using the data from one iPhone 13 is not transferable to the other two devices. 
To investigate further, the \emph{idle} activity data from the three iPhones were considered as three separate classes to see how different they are.
A Principle Component Analysis (PCA) utilising the first three components was performed on this three class mixed data and --- as shown in Figure~\ref{idle-scatter} --- it was found that the \textit{idle} activity from three iPhone 13s shows distinct patterns, making it difficult for ML to distinguish between identical components.

\begin{figure}[!t]
	\centering
	\includegraphics[width=0.3\textwidth]{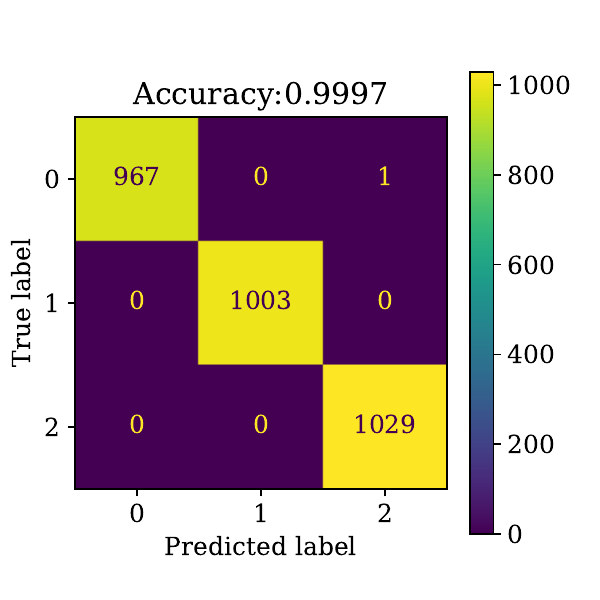}
	\caption{Confusion matrix of three iPhone 13 identical devices using an Sayakkara et al.'s EM-SCA model employing MLP machine learning approach to assess the cross-device portability of the similar type of devices.}
	\label{idle-confusion-matrix}
\end{figure}

Additionally, an MLP model was created to using the traces from the three iPhone 13s to distinguish their corresponding \emph{idle} classes, as shown in Table~\ref{MLP_idle}. 
In this MLP model, there are five hidden layers in total, as well as an output layer with three nodes that represent the idle class of three identical iPhone 13s. 
A total of 4,510,503 parameters are used to train the dataset. 
The confusion matrix of three identical iPhone 13 devices, as shown in Figure~\ref{idle-confusion-matrix}, was generated after 30 iterations.
The results indicate that the model can individually identify each \emph{idle} class of three iPhone 13 devices with 100\% accuracy. 
It reenforces the finding that a model trained using the traces from one particular device is not transferable to another device - even with the same make and model.

\begin{table}[t!]
\begin{center}
	\caption{The layout of the MLP machine learning model employing idle activities of three identical iPhone 13 dataset}
	\label{MLP_idle}
        \begin{tabular}{|l|l|l|l|}
	\hline 
	   \textbf{\footnotesize Layer (type)} & \textbf{\footnotesize Output Shape} & \textbf{\footnotesize No. of Parameters}  \\ \hline 
	   \footnotesize dense (Dense) & \footnotesize (None, 1400) & \footnotesize 2,868,600 \\ \hline 
          \footnotesize dense\_1 (Dense) & \footnotesize (None, 800) & \footnotesize 1,120,800  \\ \hline    
          \footnotesize dense\_2 (Dense) & \footnotesize (None, 500) & \footnotesize 400,500 \\ \hline 
          \footnotesize dense\_3 (Dense) & \footnotesize (None, 200) & \footnotesize 100,200 \\ \hline 
          \footnotesize dense\_4 (Dense) & \footnotesize (None, 100) & \footnotesize 20,100 \\ \hline 
          \footnotesize dense\_5 (Dense) & \footnotesize (None, 3) & \footnotesize 153 \\ \hline
        \end{tabular} 
\end{center}
\end{table}

\subsubsection{Experiment 3: Multiple Datasets from the Same Device}
\label{cross-sample-portability}


In this experiment, multiple datasets were captured from the same device, separated by time, i.e., captured on different days.
The objective was to explore how the radiation captured from the same device varies its nature across time, affecting the ability to have a stable ML model to recognise software activities running on it.
Accordingly, trace datasets were created for the iPhone 6S, iPhone 13, and iPhone 14 Pro devices repeatedly. 
Subsequently, a model was trained for each device using the dataset created from the same device at a particular day.
Then, each model was tested using the datasets of the same devices taken on another day.

\begin{figure}[!t]
	\centering
	\includegraphics[width=0.4\textwidth]{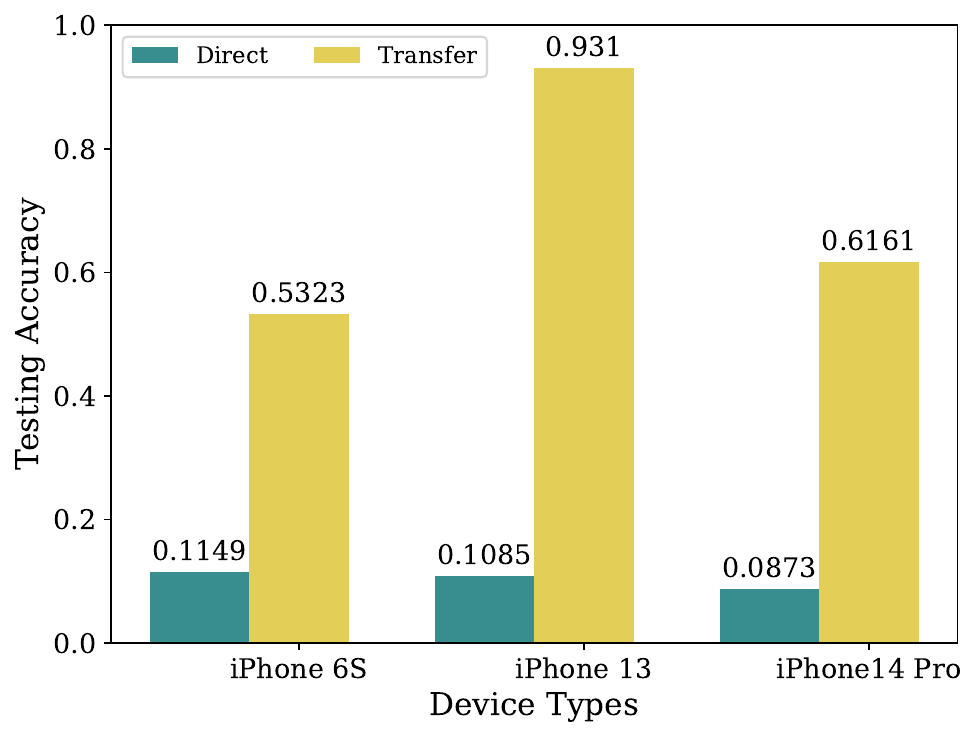}
	\caption{The accuracy of the same set of EM signals for the different iPhone 6S, 13, and 14 Pro devices was compared using the direct machine learning model of the EM-SCA approach and the transfer learning technique.}
	\label{direct_transfer}
\end{figure}

The green bars of Figure~\ref{direct_transfer} illustrate the classification accuracy of each considered iPhone device models when tested with datasets from different days.
The results indicate that traces taken across different times from the same device are considerably different.
This may have been caused due to minor variation in the data acquisition process, such as the different placement locations of the H-loop antenna, as well as variation of external EM noise sources in different days.
Under these circumstances, it may be necessary to produce EM trace datasets with a significant variety by distributing it across time.


\subsubsection{Experiment 4: Retraining Output Layer of Same Device}
\label{transfer-learning}


Instead of applying a trained model directly to a new trace dataset of the same device to validate its performance, a transfer learning technique can be used to adjust a pretrained model to a new trace dataset.
For this purpose, the final layer (i.e., output layer) of the previously trained model, is trained while the other layers are frozen across different new trace datasets as shown in the Table~\ref{MLP}. 
The previously trained model is reconstructed with 4,511,210 parameters,  among that, 1,010 trainable parameters, and  4,510,200 non-trainable parameters for training the last layer across 30 epochs. 

\begin{figure}[!t]
	\centering
	\includegraphics[width=0.4\textwidth]{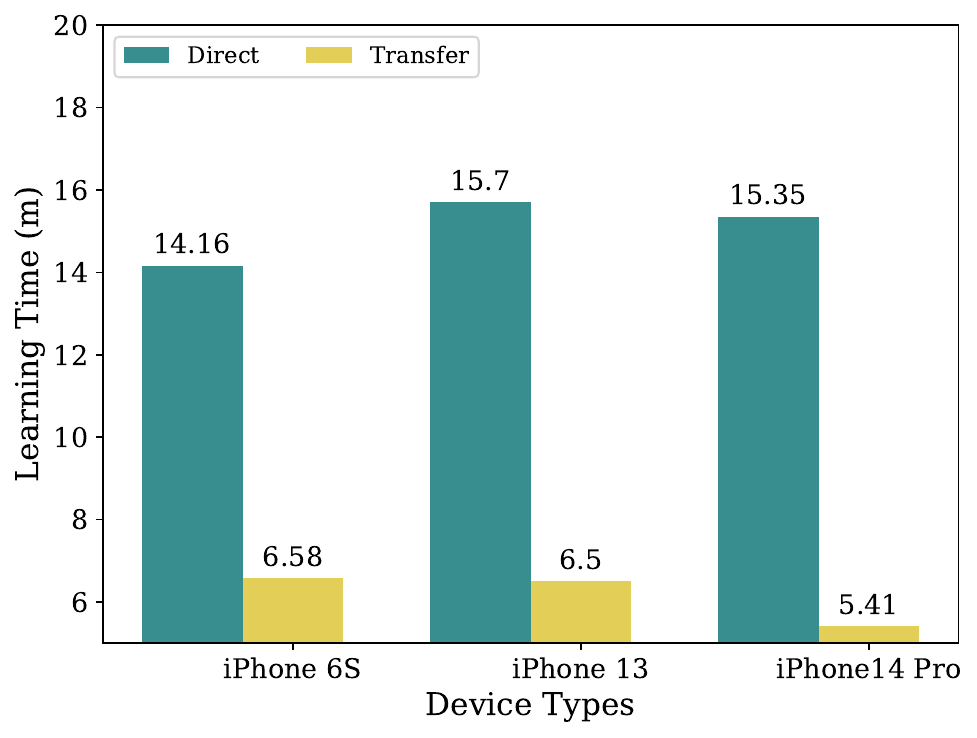}
	\caption{The learning time of the same set of EM signals for the different iPhone 6S, 13, and 14 Pro devices was compared using the direct machine learning model of the EM-SCA approach and the transfer learning technique.}
	\label{direct_transfer-time}
\end{figure}

The yellow bars in Figure~\ref{direct_transfer} illustrate the accuracy of the ML models of each iPhone models with retrained final layer by new datasets from the same device.
It is evident that the accuracy of the transfer learning models are significantly higher than models that do not have retrained output layers. 
The iPhone 13 exhibits a significant improvement over the other two versions of iPhones. 
The accuracy levels for the iPhone 6s and the 14 Pro are approximately 53\% and 61\% respectively. Therefore, the accuracy obtained using the transfer learning technique is better than that achieved using direct cross-model machine learning technique.

The training time of models were also compared between the Experiment 3 and 4; the latter is comparable to being less than it was when training a complete model initially in the former, as shown in Figure~\ref{direct_transfer-time}. 
The time required to train the entire dataset for each iPhone 6s, 13, and 14 Pro initially using the direct model for 30 epochs is displayed as green bars in Figure~\ref{direct_transfer-time}, whereas the time required to train just the final layer for 30 epochs using the transfer learning technique is shown as yellow bars. 
It is abundantly clear that using the transfer learning technique saves a significant amount of time compared to training the entire dataset, which is one of the key considerations when conducting an investigation on a smart device to obtain forensic insights.

\subsubsection{Experiment 5: Transfer Learning with Multiple Devices}
\label{CDP-transfer-learning}


Based on the findings of the previous experiments, it is evident that using a transfer learning approach improves the robustness of ML models.
To further the cross-device portability requirement, the final experiment on iPhones considered the possiblity of transferring a model across data from multiple devices of the same make and model. 
Accordingly, the pre-trained individual models of each iPhone 13 device were retrained using the trace datasets of other iPhone 13 comparable devices using transfer learning technique, by only retraining the output layer of all three iPhone 13 models.
This is done in order to validate the classification results within each individual iPhone 13 device as well as across EM radiation data across multiple iPhone 13 devices, i.e., cross-device portability of the model with the similar versions of smartphones. 
The outcomes of this experiment are illustrated in the final three columns of Table~\ref{CDP-iPhone13} under a heading \textit{Transfer learning in Sayakkara et al.'s EM-SCA approach}. 
The results highlight a significant improvement over direct pre-trained model learning on similar types of other devices. 
Despite the fact that the increment value varies, the improvement percentage is very high approximately between 60\% - 75\%.

Multiple samples from each iPhone 6S, 13, and 14 Pro were acquired in order to further evaluate the results. 
After creating customised models using the newly collected trace datasets for all the devices --- iPhone 6S, iPhone 13, and iPhone 14 Pro --- these models were used to verify the accuracy and training time while implementing the cross-model testing by using the direct models and transfer learning. 
Additionally, transfer learning was applied on all three versions of the trace datasets that were gathered while training the output layer to verify accuracy when doing cross-device, cross-model experiments on similar smartphone versions. 
The outcomes of the cross-model validations of the iPhone 6S, 13, and 14 Pro are displayed in Table~\ref{transfer-iPhone6S}, Table~\ref{transfer-iPhone13}, and Table~\ref{transfer-iPhone14Pro} respectively.

\subsection{IoT Experiment} 
\label{CDP-nordic}

\begin{figure}[!t]
	\centering
	\includegraphics[width=0.4\textwidth]{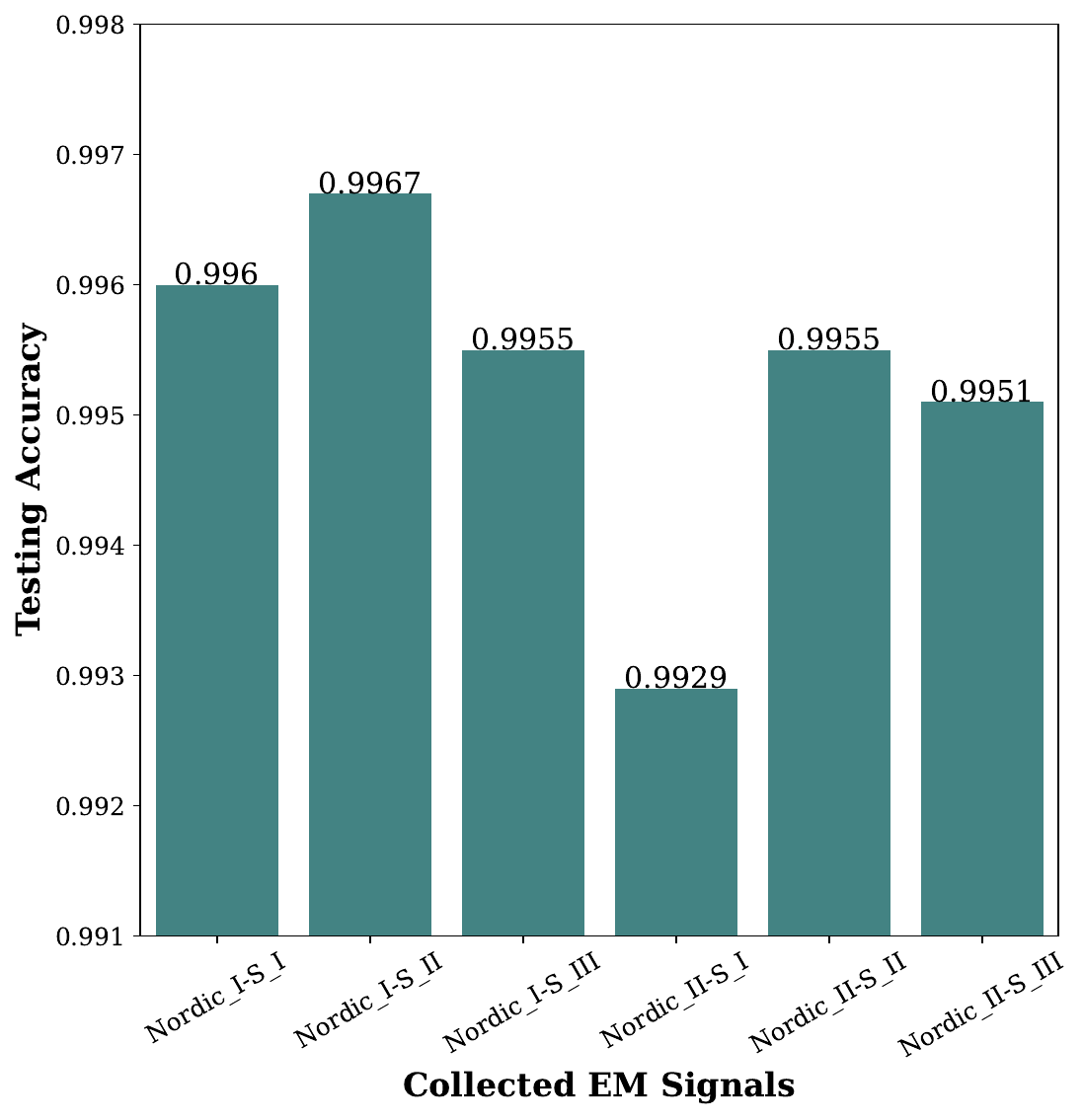}
	\caption{The individual dataset along with the EM traces for eight distinct activities for both Nordic Semiconductors are described on the x-axis, which illustrates the testing accuracy in y-axis while using the MLP machine learning model.}
	\label{nordic-Accuracy}
\end{figure}

Two identical Nordic Semiconductor nRF52-DK devices, named as \emph{Nordic-1} and \emph{Nordic-2}, were used to validate the cross-model, cross-device investigation of IoT devices, following the same procedure as the iPhone experiments. 
Eight different EM traces were captured at the 2.4 GHz system clock frequency of the nRF52-DK. 
Three sets of samples were obtained from each device, and the current EM-SCA model was used to create a tailored model for each sample of each device, as shown in Table~\ref{MLP-nordic}. 
This table illustrates the exact model that was used for the iPhone experiment, with the exception of the output layer, which is dependent on the number of activities running on the specific device. 
Additionally, the testing accuracy of the Sayakkara et al.'s EM-SCA model using MLP executed for 30 iterations in order to validate the model for each sample is shown in Figure~\ref{nordic-Accuracy}.

\begin{table}[t!]
\begin{center}
	\caption{The layout of the current machine learning model employing the newly obtained dataset from Nordic Semiconductor}
	\label{MLP-nordic}
        \begin{tabular}{|l|l|l|l|}
	\hline 
	   \textbf{\footnotesize Layer (type)} & \textbf{\footnotesize Output Shape} & \textbf{\footnotesize No. of Parameters}  \\ \hline 
	   \footnotesize dense (Dense) & \footnotesize (None, 1400) & \footnotesize 2,868,600 \\ \hline 
          \footnotesize dense\_1 (Dense) & \footnotesize (None, 800) & \footnotesize 1,120,800  \\ \hline    
          \footnotesize dense\_2 (Dense) & \footnotesize (None, 500) & \footnotesize 400,500 \\ \hline 
          \footnotesize dense\_3 (Dense) & \footnotesize (None, 200) & \footnotesize 100,200 \\ \hline 
          \footnotesize dense\_4 (Dense) & \footnotesize (None, 100) & \footnotesize 20,100 \\ \hline 
          \footnotesize dense\_5 (Dense) & \footnotesize (None, 8) & \footnotesize 808 \\ \hline 
        \end{tabular} 
\end{center}
\end{table}

Additionally, the nRF52-DK devices were used across different models and devices, in a similar fashion to the iPhone experiment. This approach aimed to evaluate the IoT device's performance by directly applying the model from one device to another and employing transfer learning techniques.
Table~\ref{transfer-nordic} displays the findings of the direct application of the model on other set of samples and the transfer learning application by training only the last layer of the pre-trained model. 

\begin{figure}[!t]
	\centering
	\includegraphics[width=0.4\textwidth]{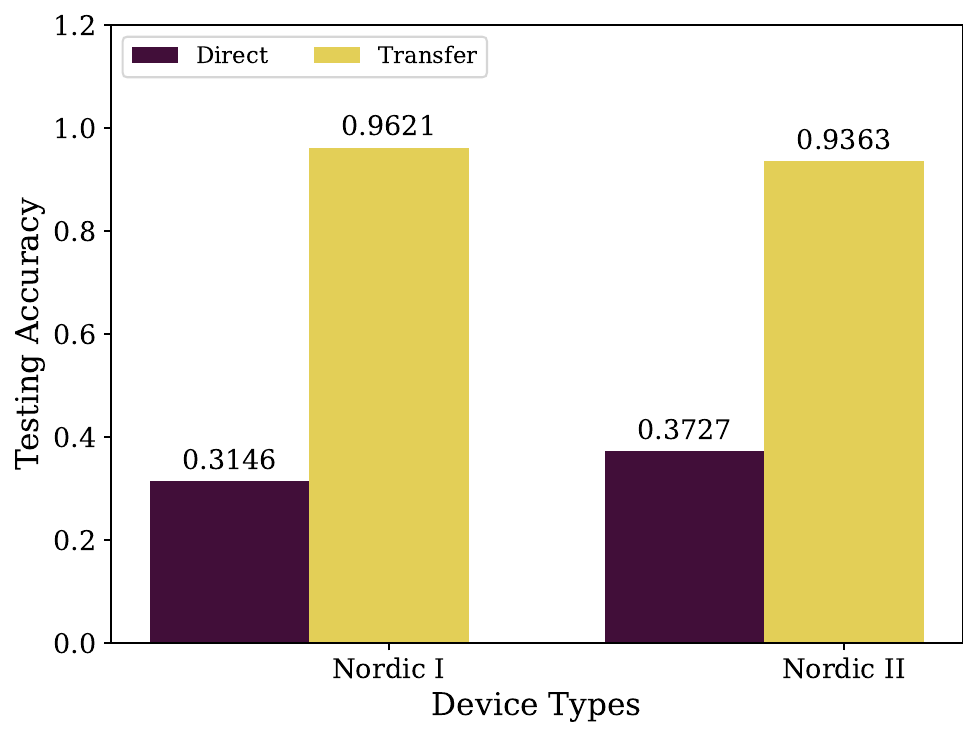}
	\caption{The accuracy of the same set of EM signals for the different nRF52-DK devices was compared using the direct machine learning model of the EM-SCA approach and the transfer learning technique.}
	\label{direct_transfer_nordic}
\end{figure}

As expected, the accuracy of the Nordic Semiconductor samples was extremely low during the cross-model, cross-device experiment, but it significantly improved after using the transfer learning technique by training the output layer to the cross-models as shown in Figure~\ref{direct_transfer_nordic}.

\section{Discussion and Future Direction}
\label{future-direction}

In a controlled environment, the EM acquisition process allows us to observe radiation without effects from external noise sources. However, real-world scenarios of digital investigations involve varying circumstances in the environment. Therefore, we have opted to capture EM radiation from devices in random locations in this study.

Noise cancellation of the raw electromagnetic (EM) traces serves as a viable strategy to enhance accuracy during the process of transfer learning for cross-device implementations. EM emissions generated by electronic devices often include unwanted background noise that can distort the integrity of the signal. This noise can arise from various sources such as electromagnetic interference, signal coupling, and environmental factors. By applying noise cancellation techniques, such as adaptive filtering or signal processing algorithms to the raw EM traces, it becomes possible to isolate the desired signal from the noise. In the context of transfer learning, where a model learns from one device and applies its knowledge to another, having accurate and consistent data is crucial.

The significance of noise cancellation lies in its ability to enhance the fidelity of the data used for training and validation. When the training data is cleaner and more representative of the true device behaviour, the resulting model is more likely to generalise well to other devices. This is particularly important for cross-device implementations, where the goal is to transfer the learned knowledge from one device to another. Overall, by employing noise cancellation techniques to refine the raw EM traces, the accuracy of transfer learning on cross-device implementations can notably be improved. This leads to more reliable and effective models that can successfully adapt knowledge across different devices, contributing to the advancement of digital forensics and related fields, e.g., security and device analysis.

In addition to noise cancellation, there are several other transfer learning techniques that can be employed to enhance the accuracy of cross-device implementations. One notable approach involves modifying the architecture of the machine learning model during the transfer learning process. Another technique is training only the input layer while keeping the rest of the layers frozen. This can be particularly effective when the lower-level features learned by the model are relevant to the new device's data. By retaining the pre-trained knowledge in the deeper layers and fine-tuning only the input layer, the model can quickly adapt to the characteristics of the new device's data, leading to improved accuracy.

Alternatively, freezing either the top or bottom part of the layers while fine-tuning the other part is another powerful technique. When the lower layers are kept frozen, the model retains the foundational features learned from the original device, while adapting its higher-level features to the new characteristics of the device. On the other hand, freezing the top layers preserves the abstract features learned from the original data, and fine-tuning the lower layers tailors the model to the specifics of the new device's data. This approach strikes a balance between reusing general features and accommodating device-specific elements.

These transfer learning techniques capitalise on the existing knowledge within a pre-trained model while enabling it to adapt to new data sources. This adaptability is particularly valuable when dealing with cross-device implementations, where data distribution and characteristics might vary significantly between devices. By incorporating these techniques, the accuracy can be significantly boosted, thereby facilitating effective knowledge transfer across different devices and ultimately enhancing the utility of the model in various applications such as classification, detection, and analysis.

In real-world use cases, a necessity to transfer a trained model using the data from a suspect device does not arise. Instead, transfer learning can be employed to generalise a model to a large number of similar devices with time. By doing so, 
the model becomes generalised enough to analyse a newly encountered device without the need for retraining using its specific data. 

\section{Conclusion} \label{conclusion}

This study unveils the challenge of reusing ML models to acquire forensic insights from smart devices, i.e. lack of cross-device portability. 
Initially, training a model directly using data from one device and testing it with data from another device yielded suboptimal outcomes.
Subsequent attempts with training models using mixed data from multiple similar devices also proved unsuccessful, highlighting the challenges involved in establishing a cohesive model across varied data sources.

Under these circumstances, the use of transfer learning strategies proved to be a crucial turning point. In particular, the adaptability of the model and performance were considerably increased by merely training the output layer. This innovation was especially noticeable in situations where various samples came from the same devices as well as identical devices. It emphasises the necessity of adaptable strategies that take into account the distinctive qualities of multiple devices while utilising transfer learning to close the gap between them. The effective use of the transfer learning technique demonstrates its potential to revolutionise EM-SCA model portability and knowledge transfer, paving the way for more precise and effective digital forensic investigations.

Despite its promise for forensic and analytical purposes, EM-SCA with transfer learning remains a complex and evolving challenge, requiring careful consideration and validation. Nevertheless, transfer learning emerges as a promising approach to advance digital forensic investigations on smart devices using EM traces. This method enhances accuracy, optimises resource utilisation, adapts to device diversity, addresses data scarcity, and enables real-time analysis by leveraging knowledge from specific or similar types of devices. As digital threats evolve, transfer learning becomes a valuable tool for forensic experts in uncovering digital evidence and securing digital environments from smart devices through EM-SCA.

\label{ending}

\section*{Appendix}

The aggregate results of the various samples performed to compare the cross-model and cross-device implementation between the iPhone 6S, 13, and 14 Pro are presented Tables~\ref{transfer-iPhone6S}, \ref{transfer-iPhone13}, and \ref{transfer-iPhone14Pro} respectively. Additionally, the results of the Nordic Semiconductor nRF52-DK are presented in Table~\ref{transfer-nordic}.

\begin{table*}[t!]
\begin{center}
	\caption{Employing empirical analysis of various samples taken at different times from the iPhone 6S to determine the testing accuracy. The samples were assessed using two methods: direct application of the present model to newly collected samples without training (referred to as "Direct"), and application of transfer learning of the existing model (referred to as "Transfer").}
	\label{transfer-iPhone6S}
        \footnotesize
        \begin{tabular}{|p{30mm}|p{20mm}|p{25mm}|p{25mm}|p{25mm}|} \hline 
            \multirow{2}{*} \textbf{\footnotesize Device Name (dataset)} & \multirow{2}{*} \textbf{\footnotesize Training Mode} & \multicolumn{3}{c|} {\textbf{\footnotesize Model Name}}  \\ \cline{3-5}
             & & \textbf{\footnotesize iPhone6S-I-Sample1} & \textbf{\footnotesize iPhone6S-I-Sample2} & \textbf{\footnotesize iPhone6S-I-Sample3}  \\ \hline 
            \multirow{2}{*} \textbf{\footnotesize iPhone6S-I-Sample1} & \footnotesize Direct & \footnotesize  0.9961 & \footnotesize  0.1214 & \footnotesize  0.1058 \\ \cline{2-5} 
                                                        & \footnotesize  Transfer & \footnotesize  - & \footnotesize  \textbf{0.5166} & \footnotesize  0.3624 \\ \hline
            \multirow{2}{*} \textbf{\footnotesize iPhone6S-I-Sample2} & \footnotesize  Direct & \footnotesize  0.1877 & \footnotesize  0.9982 & \footnotesize  0.1224 \\ \cline{2-5} 
                                                          & \footnotesize Transfer & \footnotesize \textbf{0.4795} & \footnotesize - & \footnotesize 0.3753 \\ \hline
            \multirow{2}{*} \textbf{\footnotesize iPhone6S-I-Sample3} & \footnotesize Direct & \footnotesize 0.1011 & \footnotesize 0.1186 & \footnotesize 0.9965 \\ \cline{2-5} 
                                                          & \footnotesize Transfer & \footnotesize 0.5803 & \footnotesize \textbf{0.6069} & - \\ \hline
        \end{tabular} 
\end{center}
\end{table*}

\begin{table*}[t!]
\begin{center}
	\caption{Employing empirical analysis of various samples taken at different times from the iPhone 13 to determine the testing accuracy. The samples were assessed using two methods: direct application of the present model to newly collected samples without training (referred to as "Direct"), and application of transfer learning of the existing model (referred to as "Transfer").}
	\label{transfer-iPhone13}
        \footnotesize
        \begin{tabular}{|p{27mm}|p{10mm}|p{13mm}|p{13mm}|p{13mm}|p{13mm}|p{13mm}|p{13mm}|p{10mm}|p{10mm}|} \hline 
            \multirow{2}{*} \textbf{Device Name (dataset)} & \multirow{2}{*} \textbf{Training Mode} & \multicolumn{8}{c|} {\textbf{Model Name}}  \\ \cline{3-10}
             & & \textbf{iPhone13-I-Sample1} & \textbf{iPhone13-I-Sample2} & \textbf{iPhone13-I-Sample3} & \textbf{iPhone13-I-Sample4} & \textbf{iPhone13-I-Sample5}  & \textbf{iPhone13-I-Sample6} & \textbf{iPhone13-II} & \textbf{iPhone13-III}  \\ \hline 
            \multirow{2}{*} \textbf{\footnotesize iPhone13-I-Sample1} & \footnotesize Direct & \footnotesize 0.9998 & \footnotesize 0.0690 & \footnotesize 0.0131 & \footnotesize 0.0978 & \footnotesize 0.1334 & \footnotesize 0.2507 & \footnotesize 0.0818 & \footnotesize 0.1001 \\ \cline{2-10} 
                                                        & \footnotesize Transfer & - & \textbf{\footnotesize 0.8458} & \footnotesize 0.7429 & \footnotesize 0.8223 & \footnotesize 0.7809 & \footnotesize 0.8255 & \footnotesize 0.8092 & \footnotesize 0.7400 \\ \hline
            \multirow{2}{*} \textbf{\footnotesize iPhone13-I-Sample2} & \footnotesize Direct & \footnotesize 0.1391 & \footnotesize 0.9999 & \footnotesize 0.0923 & \footnotesize 0.1957 & \footnotesize 0.1931 & \footnotesize 0.1491 & \footnotesize 0.2102 & \footnotesize 0.1000 \\ \cline{2-10} 
                                                        & \footnotesize Transfer & \footnotesize 0.8370 & \footnotesize - & \footnotesize 0.6899 & \footnotesize 0.8116 & \footnotesize 0.8068 & \footnotesize 0.7863 & \textbf{\footnotesize 0.8509} & \footnotesize 0.6868 \\ \hline
            \multirow{2}{*} \textbf{\footnotesize iPhone13-I-Sample3} & \footnotesize Direct & \footnotesize 0.0943 & 0.2478 & 0.9990 & 0.0035 & 0.3220 & 0.1212 & 0.0412 & 0.0997 \\ \cline{2-10} 
                                                        & \footnotesize Transfer & 0.8909 & \footnotesize 0.8527 & - & 0.8781 & \textbf{0.9028} & 0.8517 & 0.9008 & 0.7418 \\ \hline
            \multirow{2}{*} \textbf{\footnotesize iPhone13-I-Sample4} & \footnotesize Direct & 0.1892 & 0.0968 & 0.0000 & \footnotesize 0.9997 & 0.0042 & 0.0597 & 0.1796 & 0.0308 \\ \cline{2-10} 
                                                        & \footnotesize Transfer & 0.9262 & 0.9177 & 0.8158 & - & 0.8771 & 0.8792 & \textbf{0.9334} & 0.9194 \\ \hline
            \multirow{2}{*} \textbf{\footnotesize iPhone13-I-Sample5} & \footnotesize Direct & 0.0470 & 0.1325 & 0.0579 & 0.1644 & 0.9997 & 0.1312 & 0.1119 & 0.1000 \\ \cline{2-10} 
                                                        & \footnotesize Transfer & 0.7359 & 0.8023 & 0.7189 & 0.6657 & - & \textbf{0.8655} & 0.7844 & 0.5884 \\ \hline
            \multirow{2}{*} \textbf{\footnotesize iPhone13-I-Sample6} & \footnotesize Direct & 0.1260 & 0.1066 & 0.1200 & 0.0857 & 0.1439 & 0.9998 & 0.0857 & 0.1000 \\ \cline{2-10} 
                                                        & \footnotesize Transfer & 0.8382 & 0.7869 & 0.7318 & 0.7675 & \textbf{0.9004} & - & 0.8081 & 0.6973 \\ \hline
            \multirow{2}{*} \textbf{\footnotesize iPhone13-II} & \footnotesize Direct & 0.0612 & 0.1540 & 0.0291 & 0.1997 & 0.0708 & 0.1375 & 0.9996 & 0.0997 \\ \cline{2-10} 
                                                        & \footnotesize Transfer & 0.9572 & 0.9858 & 0.8813 & \textbf{0.9808} & 0.9601 & 0.9488 & - & 0.9024 \\ \hline
            \multirow{2}{*} \textbf{\footnotesize iPhone13-III} & \footnotesize Direct & 0.1927 & 0.1179 & 0.1041 & 0.0884 & 0.1002 & 0.0989 & 0.1139 & 0.9991 \\ \cline{2-10} 
                                                        & \footnotesize Transfer & 0.7409 & 0.7743 & 0.6533 & 0.7625 & 0.7086 & 0.6916 & \textbf{0.7784} & - \\ \hline
        \end{tabular} 
\end{center}
\end{table*}

\begin{table*}[t!]
\begin{center}
	\caption{Employing empirical analysis of various samples taken at different times from the iPhone 14 Pro to determine the testing accuracy. The samples were assessed using two methods: direct application of the present model to newly collected samples without training (referred to as "Direct"), and application of transfer learning of the existing model (referred to as "Transfer").}
	\label{transfer-iPhone14Pro}
        \footnotesize
        \begin{tabular}{|p{28mm}|p{17mm}|p{19mm}|p{19mm}|p{19mm}|p{19mm}|p{19mm}|} \hline 
            \multirow{2}{*} \textbf{Device Name (dataset)} & \multirow{2}{*} \textbf{Training Mode} & \multicolumn{5}{c|} {\textbf{Model Name}}  \\ \cline{3-7}
             & & \textbf{iPhone14Pro-I-Sample1} & \textbf{iPhone14Pro-I-Sample2} & \textbf{iPhone14Pro-I-Sample3} & \textbf{iPhone14Pro-I-Sample4} & \textbf{iPhone14pro-I-Sample5}  \\ \hline 
            \multirow{2}{*} \textbf{iPhone14Pro-I-Sample1} & Direct & 0.9962 & 0.1108 & 0.1046 & 0.0916 & 0.0999 \\ \cline{2-7} 
                                                        & Transfer & - & \textbf{0.4279} & 0.2965 & 0.3521 & 0.3239 \\ \hline
            \multirow{2}{*} \textbf{iPhone14Pro-I-Sample2} & Direct & 0.1248 & 0.9975 & 0.0969 & 0.0756 & 0.1000 \\ \cline{2-7} 
                                                         & Transfer & \textbf{0.4164} & - & 0.2967 & 0.3608 & 0.3109 \\ \hline
            \multirow{2}{*} \textbf{iPhone14Pro-I-Sample3} & Direct & 0.1100 & 0.0904 & 0.9942 & 0.1049 & 0.1000 \\ \cline{2-7} 
                                                         & Transfer & 0.1941 & 0.2065 & - & \textbf{0.2118} & 0.1796 \\ \hline
            \multirow{2}{*} \textbf{iPhone14Pro-I-Sample4} & Direct & 0.1013 & 0.1049 & 0.1027 & 0.9927 & 0.1000 \\ \cline{2-7} 
                                                         & Transfer & 0.2499 & \textbf{0.2521} & 0.2272 & - & 0.1642 \\ \hline
            \multirow{2}{*} \textbf{iPhone14Pro-I-Sample5} & Direct & 0.0873 & 0.0836 & 0.0962 & 0.1036 & 0.9990 \\ \cline{2-7} 
                                                         & Transfer & 0.6161 & \textbf{0.6336} & 0.4801 & 0.4868 & - \\ \hline
        \end{tabular} 
\end{center}
\end{table*}

\begin{table*}[t!]
\begin{center}
	\caption{Employing empirical analysis of various samples taken at different times from the Nordic Semiconductor to determine the testing accuracy. The samples were assessed using two methods: direct application of the present model to newly collected samples without training (referred to as "Direct"), and application of transfer learning of the existing model (referred to as "Transfer").}
	\label{transfer-nordic}
        \footnotesize
        \begin{tabular}{|p{27mm}|p{17mm}|p{16mm}|p{16mm}|p{16mm}|p{16mm}|p{16mm}|p{16mm}|} \hline 
            \multirow{2}{*} \textbf{Device Name (dataset)} & \multirow{2}{*} \textbf{Training Mode} & \multicolumn{6}{c|} {\textbf{Model Name}}  \\ \cline{3-8}
             & & \textbf{Nordic-I-Sample1} & \textbf{Nordic-I-Sample2} & \textbf{Nordic-I-Sample3} & \textbf{Nordic-II-Sample1} & \textbf{Nordic-II-Sample2}  & \textbf{Nordic-II-Sample3}  \\ \hline 
            \multirow{2}{*} \textbf{Nordic-I-Sample1} & Direct & 0.9960 & 0.4395 & 0.2107 & 0.2009 & 0.0043 & 0.1231 \\ \cline{2-8} 
                                                        & Transfer & - & 0.9445 & 0.9571 & 0.9004 & \textbf{0.9617} & 0.9245 \\ \hline
            \multirow{2}{*} \textbf{Nordic-I-Sample2} & Direct & 0.2472 & 0.9967 & 0.3146 & 0.2439 & 0.0447 & 0.0988 \\ \cline{2-8} 
                                                         & Transfer & 0.9389 & - & \textbf{0.9621} & 0.7728 & 0.9355 & 0.8946 \\ \hline
            \multirow{2}{*} \textbf{Nordic-I-Sample3} & Direct & 0.4606 & 0.5052 & 0.9955 & 0.1296 & 0.0025 & 0.1090 \\ \cline{2-8} 
                                                         & Transfer & \textbf{0.9470} & 0.9372 & - & 0.8132 & 0.9406 & 0.9204 \\ \hline
            \multirow{2}{*} \textbf{Nordic-II-Sample1} & Direct & 0.0057 & 0.2587 & 0.0214 & 0.9929 & 0.1263 & 0.3965 \\ \cline{2-8} 
                                                         & Transfer & 0.8075 & 0.7536 & \textbf{0.8890} & - & 0.8664 & 0.8494 \\ \hline
            \multirow{2}{*} \textbf{Nordic-II-Sample2} & Direct & 0.0006 & 0.1681 & 0.1432 & 0.1727 & 0.9955 & 0.3727 \\ \cline{2-8} 
                                                         & Transfer & 0.8795 & 0.8958 & 0.9127 & 0.9189 & - & \textbf{0.9363} \\ \hline
            \multirow{2}{*} \textbf{Nordic-II-Sample3} & Direct & 0.1282 & 0.0679 & 0.1118 & 0.1684 & 0.0712 & 0.9951 \\ \cline{2-8} 
                                                         & Transfer & 0.8795 & 0.8958 & 0.9127 & 0.9189 & \textbf{0.9363} & - \\ \hline
        \end{tabular} 
\end{center}
\end{table*}

\bibliographystyle{elsarticle-num}
\bibliography{bibfile}







\end{document}